\documentclass[aps,prl,showpacs,superscriptaddress,twocolumn]{revtex4}

\usepackage{hyperref}
\usepackage{bbm}
\usepackage{graphicx}
\usepackage{amsmath}
\usepackage{amssymb}
\usepackage{mathscinet}
\usepackage{amsfonts,amsmath}
\usepackage{color}

\newcommand{\id}{I}
\def\ket#1{| #1 \rangle}
\def\bra#1{\langle #1 |}

\newcommand{\tr}[1]{\text{Tr} #1}

\newcommand{\m}{\mathcal }

\newtheorem{theorem}{Theorem}

\newtheorem{lemma}[theorem]{Lemma}

\begin{document}

\title{Private Quantum Subsystems}

\author{Tomas Jochym-O'Connor}

\affiliation{Institute for Quantum Computing, University of Waterloo, Waterloo, Ontario, N2L 3G1, Canada}
\affiliation{Department of Physics \& Astronomy, University of Waterloo, Waterloo, Ontario, N2L 3G1, Canada}

\author{David W. Kribs}

\affiliation{Institute for Quantum Computing, University of Waterloo, Waterloo, Ontario, N2L 3G1, Canada}
\affiliation{Department of Mathematics \& Statistics, University of Guelph, Guelph, Ontario, N1G 2W1, Canada}

\author{Raymond Laflamme}

\affiliation{Institute for Quantum Computing, University of Waterloo, Waterloo, Ontario, N2L 3G1, Canada}
\affiliation{Department of Physics \& Astronomy, University of Waterloo, Waterloo, Ontario, N2L 3G1, Canada}
\affiliation{Perimeter Institute for Theoretical Physics, 31 Caroline Street North, Waterloo, Ontario, N2L 2Y5, Canada}

\author{Sarah Plosker}

\affiliation{Department of Mathematics \& Statistics, University of Guelph, Guelph, Ontario, N1G 2W1, Canada}

\date{\today}

\begin{abstract}
We investigate the most general notion of a private quantum code, which involves the encoding of qubits into quantum subsystems and subspaces. We contribute to the structure theory for private quantum codes by deriving testable conditions for private quantum subsystems in terms of Kraus operators for channels; establishing an analogue of the Knill-Laflamme conditions in this setting. For a large class of naturally arising quantum channels, we show that private subsystems can exist even in the absence of private subspaces. In doing so, we also discover the first examples of private subsystems that are not complemented by operator quantum error correcting codes; implying that the complementarity of private codes and quantum error correcting codes fails for the general notion of private quantum subsystem.
\end{abstract}


\maketitle

\section{Introduction \& Background}

The most essential primitive for private communication between two parties, Alice and Bob, in classical computation is the one-time pad. In such a scheme, the two parties share a secret key that is unknown to an external observer Eve; this key enables reliable communication by the parties as the message appears to be a random mixture of input bits from Eve's viewpoint without the key.

Private quantum codes were initially introduced as the quantum analogue of the classical one-time pad. The basic setting for a ``private quantum channel'' \cite{AMTW00,BR03} is as follows: Alice and Bob share a private classical key that Alice uses to inform Bob which of a set of unitary operators $\{ U_i \}$ she has used to encode her quantum state: $\rho \mapsto U_i \rho U_i^\dagger$. With this information in hand, Bob can decode and recover the state $\rho$ without disturbing it. The set of unitaries $\{U_i\}$ and the probability distribution $\{p_i\}$ that makes up the random key which determines the encoding unitary are shared publicly. Thus, without further information, Eve's description of the system is given by the random unitary channel $\Phi(\rho) = \sum_i p_i U_i \rho U_i^\dagger$. By selecting certain sets of unitary operators with appropriate coefficients, the random unitary channel will provide Eve with no information about the input state.

The body of work on private quantum codes now includes a variety of other applications, with realizations both as subspaces and subsystems of Hilbert space. Private shared reference frames exploit private subspaces and subsystems that also arise from the ignorance associated with an eavesdropper's description of a system  \cite{BRS04,BHS05}. The notion of using mixed state ancilla qubits to encode information, which can be viewed as subsystem encodings, has also been studied in the context of quantum secret sharing~\cite{CGL99,CGS02}. There, the goal is to encode information into a globally mixed state of $n$~qubits such that to recover the quantum information one would need access to $k$ qubits of the global state, where any fewer would yield no information regarding the initial state. Using mixed states allows for the increase of $k$ for a fixed $n$, thus solidifying the idea that mixed state encodings increase privacy. There are also bridges between these works and quantum error correction, formalized by the complementarity results of \cite{KKS08}. Connections between the study of private quantum codes and the theory of operator algebras have recently been found as well \cite{CKPP12}.


In this Letter we consider the most general notion of a private quantum code \cite{AMTW00,BR03,BRS04}, which involves the encoding of quantum bits into subsystems. Private quantum channels, private subspaces, and what we refer to as ``operator'' private subsystems---are captured as special cases of this general phenomena. We consider a class of phase damping channels throughout the presentation that highlights the main differences between mappings on subsystems and subspaces. Most surprisingly, we show that certain classes of channels can only be private in the subtle subsystem sense; thus establishing that private subsystems can exist in the absence of private subspaces.

We also make the first significant move toward a structure theory for private quantum codes; specifically we set out algebraic conditions that characterize privacy of a code in terms of the Kraus operators for a given quantum channel. This can be viewed as an analogue of the set of Knill-Laflamme conditions \cite{KL97} from quantum error correction to this setting, and indeed we discuss further connections with error correction. In particular we show that complementarity of private and error-correcting codes fails at the most general level, and we point out a potentially new type of quantum error-correcting code.

We now describe our notation and nomenclature. Given a quantum system $S$, with (finite-dimensional and complex) Hilbert space also denoted by $S$, we will use customary notation such as $\rho$, $\sigma$ for density operators. The set of linear operators on $S$ is denoted by $\mathcal{L}(S)$. Linear maps on $\mathcal{L}(S)$ can be viewed as operators acting on the operator space $\mathcal{L}(S)$. We use the term \emph{(quantum) channel} to refer to a completely positive and trace-preserving linear map $\Phi:\mathcal{L}(S)\rightarrow \mathcal{L}(S)$. Such maps describe (discrete) time evolution of open quantum systems in the Schr\"{o}dinger picture, and can always be written in the Choi-Kraus operator-sum form $\Phi(\rho) = \sum_i V_i \rho V_i^\dagger$ for some operators $V_i$ in $\mathcal{L}(S)$ satisfying $\sum_iV_i^\dagger V_i=I$. The composition of two maps will be denoted by $\Phi\circ \Psi(\sigma )=\Phi(%
\Psi(\sigma ))$.

A (linearly closed) subspace $C$ of $S$ is said to be a \emph{private subspace} for $\Phi$ if
there is a density operator $\rho_0$ on $S$ such that
$\Phi (\ket{\psi}\bra{\psi}) = \rho_0$ for all pure states $\ket{\psi}$ in $C$. By linearity, $\Phi (\rho)
= \rho_0$ for all $\rho$ in $\mathcal L(C)$. We could also consider a collection of private states not associated with a subspace of the Hilbert space, but, as in quantum error correction, we wish to allow for arbitrary superpositions of our code states and this demands the set of states considered are linearly closed.

A quantum system $B$ is a \textit{subsystem} of $S$ if we can write
$S=(A\otimes B)\oplus (A\otimes B)^{\perp }$.
This definition is symmetric in that $A$ is also considered a subsystem of $S$. The sub\emph{spaces} of $S$ can
be viewed as subsystems $B$ for which $A$ is one-dimensional. A subscript such as $\sigma_B$ means the operator belongs to $\mathcal{L}(B)$. A subsystem $B$ is a \emph{private subsystem} for $\Phi$ if there is a $\rho_0 \in \mathcal{L}(S)$ and $\sigma_A \in \mathcal{L}(A)$ such that
\begin{eqnarray}\label{private_definition}
\Phi (\sigma_A \otimes \sigma_B) = \rho_0 \quad \forall \sigma_B\in \mathcal{L}(B).
\end{eqnarray}
The case of random unitary channels $\Phi$ in
Eq.~(\ref{private_definition}) was first considered in
\cite{AMTW00,BR03} where the terminology \emph{private quantum channels} was used, and the case of general channels $\Phi$ was
formalized in \cite{BRS04} where private subsystems were given the extra prefix ``completely'' that we have dropped. If Eq.~(\ref{private_definition}) holds for all $\sigma_A$, as opposed to a single state
$\sigma_A$, then we shall refer to $B$ as an \emph{operator private subsystem} (since these are precisely the private subsystems that are complementary to operator quantum error-correcting subsystems discussed below).

\section{Private Subsystems In The Absence Of Private Subspaces}

An operator private subsystem is one in which the private channel
splits into a product of maps on the individual subsystems $A$ and $B$ when
the channel is restricted to the combined product subspace $A \otimes B$. Such private subsystems cannot exist
without the existence of private subspaces; indeed, if Eq.~(\ref{private_definition})
holds for all states on $A$, it follows that every sub\emph{space}
$\ket{\psi}\otimes B$ is private for $\Phi$ for any fixed pure state $\ket{\psi}$ on $A$.

Even though the definition given by
Eq.~(\ref{private_definition}) allows for the possibility of
examples of private subsystems that do not extend to private subspaces, the private subsystems
exhibited in the literature
\cite{AMTW00,BR03,BRS04,BHS05} thus far have either been of operator type,
or are already subspaces. Here we present the first examples of private
subsystems for which there are no private subspaces that exist; in particular these are private subsystems that are
not of operator type. Our motivating class of channels is
built upon a simple phase damping model. We begin the discussion
by recalling the most basic private quantum channel and asking
some basic questions on quantum privacy.


The completely depolarizing channel ($\Phi(\rho) = \frac{1}{\dim
S}\id$ for all $\rho$) is an easy to describe example of a
quantum channel that is private. In this case the entire Hilbert
space acts as a private code for the channel, and so in order to
implement such a private channel a full set of Pauli rotations
must be available. This leads to a very basic question in the
study of private quantum codes: Do there exist channels with fewer
physical operations such that we can still encode qubits for
privacy?

Perhaps the simplest class of channels one could imagine would
be the family of phase damping channels that can be applied to any
qubit of a larger Hilbert space~$\m{S}$ of $n$~qubits,
\begin{eqnarray}
\Lambda_i (\rho) = \frac{1}{2}(\rho + Z_i \rho Z_i ), \qquad \forall \rho \in \m{L}(S).
\end{eqnarray}
A single qubit phase damping channel is not private. Yet we can
ask: can composing the phase damping channel on multiple qubits
yield a private subspace~$C \subseteq\m{S}$? Such a question is
analogous to the sort of questions that have been asked in quantum
error correction for some time; for example, given a set of errors
that are uncorrectable on a single qubit, does there exist a
larger Hilbert space such that the action of the error on the
encoded Hilbert space is correctable? The answer to such a
question in quantum error correction is yes, as demonstrated by
the five-qubit code which corrects for arbitrary single-qubit
errors, an error that would be uncorrectable if one did not have
access to a larger Hilbert space to encode the quantum information
into a quantum code.

We shall define the map $\Lambda$ as the composition of the maps $\Lambda_i$ on each of the $n$~qubits of the state $\rho \in  S$,
\begin{eqnarray}
\Lambda (\rho) &= \Lambda_n \circ \Lambda_{n-1} \circ \dots \circ \Lambda_1 (\rho).
\end{eqnarray}
Equivalently one could consider the $n$-product map
$\Lambda_1^{\otimes n}$ of the single qubit channel $\Lambda_1$.
For any input state $\rho$, this channel will decohere all
off-diagonal terms in the computational basis; as such, the
resulting output density matrix will be diagonal.

Consider the case when $n=2$. Every output state of $\Lambda$ has the form
\begin{eqnarray}
\label{eq:outputstate}
\rho_0 = \frac{1}{4}\Big( II  + \alpha IZ + \beta ZI + \gamma ZZ \Big),
\end{eqnarray}
where $I$ and $Z$ are the one-qubit identity and Pauli $Z$
matrices. The goal is to find a subspace $C$ of dimension~2 and a
state $\rho_0 \in \m{L}(S)$ such that $\Lambda(\rho) = \rho_0$
$\forall \rho \in \m{L}(C)$. This would show that $\Lambda$ has a
private qubit subspace, defined by a pair of orthogonal logical states
$\ket{0_L}$, $\ket{1_L}$ in $C$.

However, one can show that such a subspace \emph{does not} exist.
In fact we can prove the following more general result, which
applies to the channels $\Lambda$ directly, and can be extended to channels
with commuting normal Kraus operators as well. We leave the proof for \cite{JKLP12}.

\begin{lemma}
Let $\Phi$ be a random unitary channel with mutually commuting Kraus operators.
Then $\Phi$ has no private subspaces.
\label{lem:RandomUnitaryChannel}
\end{lemma}

Is this the end of the story? This result is intuitive---at first
glance it certainly does not ``feel'' as though we should be able
to find private codes for channels such as the phase damping maps
$\Lambda = \Lambda_n \circ \cdots \circ \Lambda_2 \circ
\Lambda_1$ due to the preservation of information stored in
the diagonal elements of the initial density matrix.
Moreover, the experience with operator private
subsystems, which demand the existence of private subspaces, also
suggests we can go no further with these channels. Somewhat
surprisingly, we do find private subsystems for these channels,
and necessarily they are not of the type exhibited before.

Indeed, consider the following logically encoded qubits in two-qubit
Hilbert space:
\begin{eqnarray}\label{subsystem_encoding}
\rho_L=\frac14(II+\alpha XX + \beta YI+\gamma ZX).
\end{eqnarray}
This describes a single qubit encoding, as
Eq.~(\ref{subsystem_encoding}) describes the coordinates for a
logical Bloch sphere in two-qubit Hilbert space with logical Pauli
operators given by $X_L=XX, Y_L=YI, Z_L=ZX$. Now, observe that the dephasing map $\Lambda = \Lambda_2 \circ
\Lambda_1$ acting on each density operator $\rho_L$ produces an
output state that is maximally mixed; that is, $\Lambda(\rho_L) =
\frac14 \, II$ for all $\rho_L$. Thus, we see that
Eq.~(\ref{subsystem_encoding}) yields a private two-qubit code for the
dephasing map $\Lambda$. However, we know from
Lemma~\ref{lem:RandomUnitaryChannel} that the input space cannot
be a subspace, and we have already noted this implies it also
cannot be an operator subsystem. It is however still a private
subsystem in the sense of Eq.~(\ref{private_definition}). Let us discuss the encoding in more detail.

\begin{figure}[h]
\centering
\includegraphics[width=0.30\textwidth]{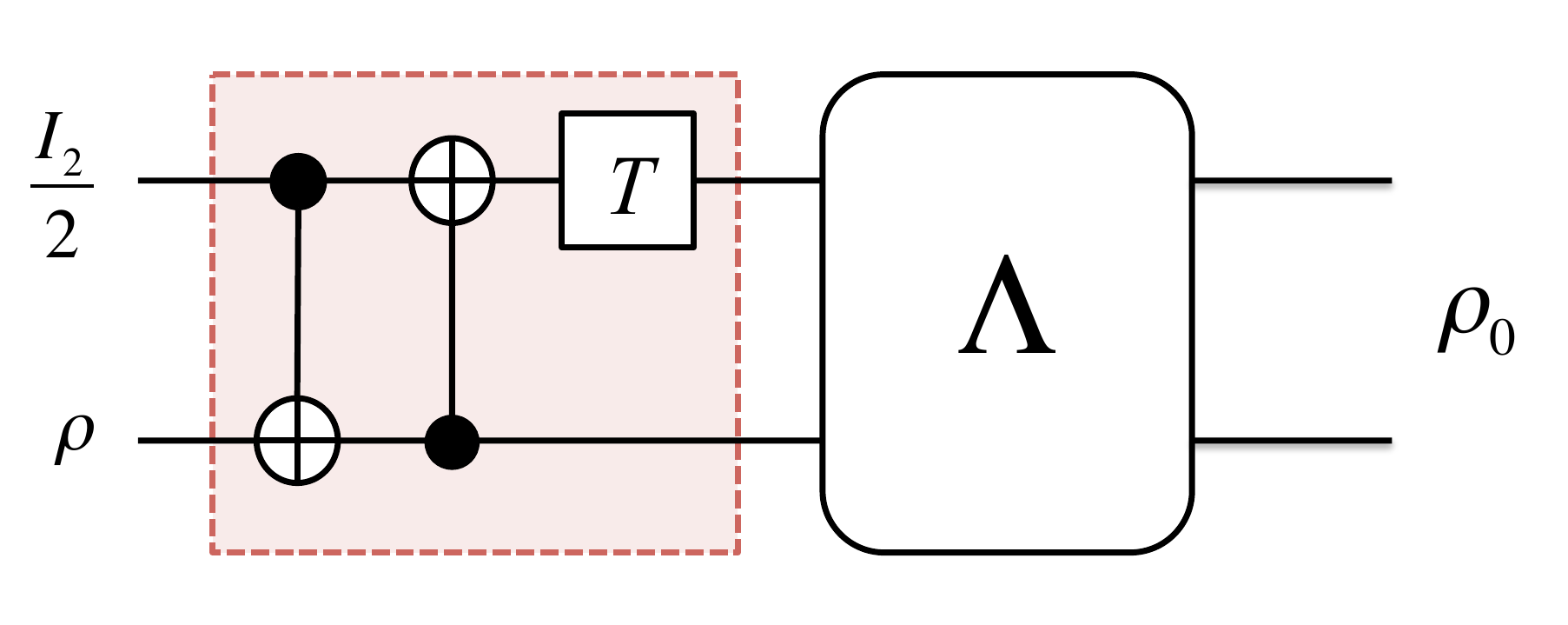}
\caption{The gates in the red box implement the encoding of an arbitrary two-qubit state belonging to the $I \otimes \mathbb{C}^{2\times 2}$ algebra into encoded states of the form of Eq.~\eqref{subsystem_encoding}. The encoded two-qubit state subjected to the two-qubit phase damping channel $\Lambda~=~\Lambda_2\circ~\Lambda_1$ results in an output of the completely mixed state on two qubits, $\rho_0 = \frac{1}{4} II$.}
\label{fig:PrivateChannel}
\end{figure}

The logical encoding of a single qubit into a two-qubit subsystem is shown by the unitary operation given by the red boxed region in Figure~\ref{fig:PrivateChannel}. The mapping, given by a pair of CNOT gates and the $T = \frac1{\sqrt{2}} (\ket{0}(\bra{0} + \bra{1}) + i \ket{1}(\bra{0} - \bra{1}))$, shows a unitary equivalency between the two-qubit operator algebra~$I_2 \otimes \mathbb{C}^{2\times 2}$ and the encoded logical qubit through the following transformation of the basis elements of $\mathbb{C}^{2\times 2}$,
\begin{eqnarray*}
IX \longmapsto  & IX &\longmapsto   XX \longmapsto ZX\\
IY \longmapsto  & ZY  &\longmapsto   YX \longmapsto XX\\
IZ \longmapsto  & ZZ &\longmapsto    ZI  \longmapsto IY.
\end{eqnarray*}
More generally, a logical qubit encoding into a subsystem of a $n$-qubit Hilbert space can be constructed to privatize the $n$-qubit phase damping channel $\Lambda = \Lambda_n \circ \cdots \circ \Lambda_2 \circ \Lambda_1$, which by Lemma~\ref{lem:RandomUnitaryChannel} cannot have a private subspace.

\begin{theorem}
For any $n$-qubit Hilbert space $\mathcal{H}$, there exist quantum channels $\Phi$ for which a private quantum subsystem $B$ of $\mathcal{H}$ can be constructed in the absence of the existence of any private quantum subspace $C \subseteq \mathcal{H}$.
\label{thm:RandomUnitaryChannel}
\end{theorem}

\section{Testable Conditions For Private Quantum Codes}

If we are given a quantum channel $\Phi(\rho) = \sum_i V_i \rho V_i^\dagger$ and a subsystem $B$, we can ask if it is possible to decide whether $B$ is private for $\Phi$; and more to the point, we can ask if this can be answered in terms of the Kraus operators $V_i$ for the channel. The analogous question in quantum error correction is answered by the fundamental Knill-Laflamme conditions \cite{KL97}, which provide an explicit set of algebraic constraints in terms of the Kraus operators and the code, and allow one to test whether a given code is correctable for a channel. The generalization of these conditions to the case of operator error-correcting subsystems was established in \cite{KLP05,KLPL05,NiPo07}.

The following result answers this question for private quantum subsystems. In addition to Kraus operators, we would expect the algebra to include the fixed $A$ state $\sigma_A$ and output state $\rho_0$---observe that this information is indeed included in the conditions.

\begin{theorem}
 A subsystem $B$ is private for a channel
$\Phi(\rho) = \sum_i V_i \rho V_i^\dagger$ with fixed $A$
state $\sigma_A$ and output state $\rho_0$ if and only if
there are complex scalars $\lambda_{ijkl}$ forming an isometry matrix $\lambda = (\lambda_{ijkl})$ such that
$\sqrt{p_k }V_j\ket{\psi_{A,k}} = \sum_{i,l}\lambda_{ijkl} \sqrt{q_l}\ket{\phi_{l}}\bra{\psi_{B,i}}$,
where $\ket{\psi_{A,k}}$ ($p_k$) and $\ket{\phi_{l}}$ ($q_l$) are eigenstates (eigenvalues) of
$\sigma_A$ and $\rho_0$ respectively, $\ket{\psi_{B,i}}$ is an orthonormal basis
for $B$, and where $\ket{\psi_{A,k}}$ is viewed as a channel from $B$ into $S$.
\end{theorem}

The key observation in establishing this result is that the left and right hand sides of Eq.~(\ref{private_definition}) each define channels from $B$ to $S$ which are in fact the same. One can then use basic results from the theory of completely positive maps to obtain the equations spelled out in the theorem. More details on the theory will be presented in \cite{JKLP12}.

It is important to note that this result is new even for private sub\emph{spaces}. In the notation of the theorem for that case, $A$ is one-dimensional and $B$ is the subspace. If we let $P_B$ be the projector of $S$ onto $B$, then we see that the characterization of privacy is given by the conditions: $V_j P_B = \sum_{i,l}\lambda_{ijl} \sqrt{q_l}\ket{\phi_{l}}\bra{\psi_{B,i}}$ for all $j$. As a simple illustration, in the case of the completely depolarizing channel on $N$-dimensional Hilbert space, $P_B$ is the identity operator and these conditions reduce to the Kraus operators satisfying $\sqrt{N}\, V_j = \sum_{i_1,i_2}  \lambda_{i_1i_2 j}  \ket{i_1}\bra{i_2}$, for some choice of orthonormal bases $\ket{i_1}$ and $\ket{i_2}$ and unitary matrix $(\lambda_{i_1i_2 j})_{i_1,i_2}$. One can also phrase the private subspace conditions neatly via the Heisenberg picture in terms of the dual map $\Phi^\dagger$, which has Kraus operators $V_j^\dagger$, as follows~\cite{KP12}: $P_B \Phi^\dagger (M) P_B = \tr(M\rho_0) P_B$ for all (arbitrary) observables $M$.

Of course the theorem applies to general private subsystems as well. Here we point out how the 2-qubit phase damping channel $\Lambda$ can be assembled from this result. In that case both $A$ and $B$ are spanned by $\{\ket{0}$, $\ket{1}\}$. The eigenstates of $\rho_0=\frac14 I_4$ are $\{\ket{00}, \ket{01}, \ket{10}, \ket{11}\}$, each having eigenvalue $\frac14$. The eigenstates of $\sigma_A=\frac12 I_2$ are $\{\ket{\psi_{A,k}}\}=\{\ket{0}, \ket{1}\}$, with corresponding eigenvalues $\frac12$. For brevity, we omit the matrix calculations for the operators $V_j\ket{\psi_{A,k}}$ here; we simply note that they are $4\times 2$ matrices formed with $2\times 2$ Pauli operators and zero blocks. The scalar-valued matrix $\lambda=(\lambda_{ijkl})$ is indeed an isometry. Furthermore, because the number of  operators $V_j\ket{\psi_{A,k}}$ agrees with the number of operators $\ket{\phi_{l}}\bra{\psi_{B,i}}$ (namely, 8), the matrix $\lambda$ is in fact unitary.

\subsection{Complementarity and Quantum Error Correction} Several links have been made between quantum error correction and  quantum privacy. In the case of operator private subsystems and operator error-correcting subsystems, the complementarity theorem of \cite{KKS08} discussed below established an algebraic bridge between the two subjects. This firmly links the operator quantum error correction theory to that of operator private subsystems---results in one field can immediately be exported to the other. Thus, it is natural to ask whether such a result holds in this more general setting. To answer this we need the concept of complementary channels.

As a consequence of the Stinespring dilation theorem, every channel $\Phi$ may be seen to arise from an environment Hilbert space $E$, a pure state $|\psi \rangle $ on the environment, and a unitary operator $U$ on the composite $%
S\otimes E $ in the following sense: $\Phi(\rho )=\tr_{E}\big(%
U(\rho \otimes \ket{\psi}\bra{\psi})U^\dagger\big).$ Tracing out the system instead yields a complementary channel: $\Phi^{\sharp}(\rho )=\tr_{S}\big(U(\rho \otimes
\ket{\psi}\bra{\psi})U^\dagger\big).$ The uniqueness built into the theorem yields a certain uniqueness for such a pair of channels, so that we talk of ``the'' complementary channel $\Phi^\sharp$ for a given channel $\Phi$ \cite{Hol06,KMNR07}.

We have already discussed operator private subsystems---the essential difference being that instead of a single state on $A$, it is demanded that Eq.~(\ref{private_definition}) holds for all states on $A$. Similarly, an operator quantum error-correcting subsystem $B$ for a channel $\mathcal E$ \cite{KLP05,KLPL05} requires the existence of a correction operation $\mathcal R$ such that: $\forall \sigma_A$ $\forall \sigma_B$, $\exists \tau_A$ for which $\mathcal R \circ \mathcal E (\sigma_A \otimes \sigma_B) = \tau_A \otimes \sigma_B$. The main result of \cite{KKS08} shows that $B$ is private for $\Phi$ if and only if it is error-correcting for $\Phi^\sharp$.

Does this result extend to the more general setting? The Kraus operators of the complementary map~$\Lambda^{\sharp}$ are four orthogonal rank-one projectors in two-qubit Hilbert space, and in particular the map determines a von Neumann measurement. No error-correcting subsystem can be extracted in such a setting; moreover, when the input space is restricted to be that of our example, the complementary map is private. Thus, not only does the complementarity result fail, it fails dramatically. We save these calculations and further analysis for \cite{JKLP12}.

This discussion motivates the following observation: The notion of an operator quantum error-correcting subsystem can be expanded to mimic the general definition of a private quantum subsystem. Indeed, a revised definition analogous to that of Eq.~(\ref{private_definition}) could be proposed as follows: $B$ is \emph{correctable} for $\mathcal E$ if there exists an operation $\mathcal R$ such that for all $\sigma_B$ and some \emph{fixed} states $\sigma_A,\tau_A$, we have $\mathcal R \circ \mathcal E (\sigma_A \otimes \sigma_B) = \tau_A \otimes \sigma_B.$ This is a potentially new notion of quantum error-correcting code.


\section{Conclusion \& Outlook}

We have studied the most general notion of private quantum subsystems. Taking motivation from quantum error correction, specifically the Knill-Laflamme conditions, we presented algebraic conditions that characterize when a code is private for a given channel. This result is new even for private subspaces, and opens up questions such as: is there an analogue of the stabilizer formalism from quantum error correction for quantum privacy? We analyzed the development of private subsystems for the special case given by the composition of phase damping channels on many qubit Hilbert~spaces. While each individual channel of this form is not private, the composition of such channels were shown to contain a private single qubit subsystem. Yet, for such channels, and for a wide class of more general channels, no private subspace or operator private subsystem exists. Moreover, we discussed how the channel fails to have the corresponding complementary error-correctable pair as in the case of operator subsystems. Nevertheless, this analysis naturally led us to define a potentially new form of quantum error-correcting code and warrants, along with the topic of channel complementarity, further investigation. Finally, preliminary discussions also suggest there may be yet unexplored connections between the study of private quantum subsystems and privacy in classical communication. We will continue and expand on the work initiated here elsewhere.

\section{Acknowledgments}

We are grateful to Robert Spekkens for an interesting discussion, and we thank the referees for helpful comments on our initial submission. T.~J.-O.\ was supported by an Ontario Graduate Scholarship. D.W.K.\ was supported by NSERC. R.L.\ was supported by NSERC, CIFAR, and Industry Canada. S.P.\ was supported by an NSERC Graduate Scholarship.

\end{document}